\documentclass{emulateapj}
\submitted{ \today}
\journalinfo{Accepted in ApJ January 17, 2019}
\usepackage{natbib}
\usepackage{xcolor}
\usepackage{color}
\usepackage{comment}
\usepackage{amsmath}
\definecolor{color}{RGB}{25,25,112}
\definecolor{negro}{RGB}{0,0,0}
\definecolor{colorurl}{RGB}{25,25,112}

\usepackage[hyperfootnotes=false]{hyperref}
\hypersetup{
  colorlinks,
  citecolor=color,
  linkcolor=color,
  urlcolor=colorurl}
 
\newcommand{\be}{\begin{equation}}
\newcommand{\ee}{\end{equation}}
\newcommand{\bse}{\begin{subequations}}
\newcommand{\ese}{\end{subequations}}
\newcommand{\bary}{\begin{eqnarray}}
\newcommand{\eary}{\end{eqnarray}}

\shorttitle{Early Photometry of GRB 180205A}
\shortauthors{Becerra et al.}

\begin{document}

\title{Late Central Engine Activity in GRB 180205A}
 \author{%
Becerra,~R.~L.$^{1}$;
Watson,~A.~M.$^{1}$;
Fraija,~N.$^{1}$;
Butler,~N.~R.$^{2}$;
Lee,~W.~H.$^{1}$;
Troja,~E.$^{3,4}$;
Rom\'an-Z\'u\~niga,~C.~G.$^{5}$;
Kutyrev,~A.~S.$^{3,4}$;
\'Alvarez~Nu\~nez,~L.~C.$^{1}$;
\'Angeles,~F.$^{1}$;
Chapa,~O.$^{1}$; 
Cuevas,~S.$^{1}$;
Farah,~A.~S.$^{1}$; 
Fuentes-Fern\'andez,~J.$^{1}$;
Figueroa,~L.$^{5}$;
Langarica,~R.$^{1}$;
Quir\'os,~F.$^{5}$;
Ru\'iz-D\'iaz-Soto,~J.$^{1}$;
Tejada,~C.~G.$^{5}$;
Tinoco,~S.~J.$^{1}$;
}

\address{$^1$ Instituto de Astronom{\'\i}a, Universidad Nacional Aut\'onoma de M\'exico, Apartado Postal 70-264, 04510 M\'exico, CDMX, M\'exico;\\
$^2$ School of Earth and Space Exploration, Arizona State University, Tempe, AZ 85287, USA;\\
$^3$ Department of Astronomy, University of Maryland, College Park, MD 20742-4111, USA;\\
$^4$ Astrophysics  Science  Division,  NASA  Goddard Space Flight Center, 8800 Greenbelt  Rd, Greenbelt, MD 20771, USA;\\
$^5$  Instituto de Astronom{\'\i}a, Universidad Nacional Aut\'onoma de M\'exico, Unidad Acad\'emica en Ensenada, 22860 Ensenada, BC, Mexico.
}

\begin{abstract}
\noindent
We present optical photometry of the afterglow of the long GRB 180205A with the COATLI telescope from 217 seconds to about 5 days after the {\itshape Swift}/BAT trigger. We analyse this photometry in the conjunction with the X-ray light curve from {\itshape Swift}/XRT.
The late-time light curves and spectra are consistent with the standard forward-shock scenario.
However, the early-time optical and X-ray light curves show non-typical behavior; the optical light curve exhibits a flat plateau while the X-ray light curve shows a flare.    
We explore several scenarios and conclude that the most likely explanation for the early behavior is late activity of the central engine.

\end{abstract}
\begin{center}
\end{center}

\section{Introduction}

Gamma-ray bursts (GRBs) are, during their brief lives, the most luminous events in the Universe. Observationally, they are typically classified according to their duration $T_{90}$ \citep{Kouveliotou93} as short GRBs and long GRBs. Short GRBs are thought to arise from the fusion of compact objects \citep{ls76,bp86,bp91,2007NJPh....9...17L}, whereas long GRBs are thought to arise from core-collapse supernovae in massive stars \citep{1993ApJ...3505..273W,mw99}.

The most successful theory for interpreting the electromagnetic radiation from both types of GRBs is the fireball model \citep[recently reviewed by ][]{2015PhR...561....1K}. This model distinguishes two stages: the prompt or early phase and the afterglow or late phase. The prompt phase is simultaneous with bright emission in gamma rays produced by internal shocks in the jet driven by the central engine. The afterglow phase is produced by the external shocks between the outflow and the circumstellar environment \cite[e.g.,][]{2000APJ...532....286K,  2015ApJ...804..105F}. While the general ideas behind the fireball model seem to be clear, it is important to compare the detailed predictions of the model with observational data of GRBs in order to determine which facets of the model are important in individual case, to determine the ranges of macrophysical and microphyical parameters, and to understand the diversity of GRBs. This process leads to a better understanding both of GRBs and of the environments in which they develop.

These ideas are the main motivation of this paper. Here, we present photometric observations of the long GRB 180205A in the optical and in X-rays during the afterglow phase and compare them to the detailed predictions of the fireball model.

GRB 180205A was one of the brightest GRBs in the last few years and presented a bright optical counterpart. It was well observed by many collaborations; in total, there were 22 GCN Circulars related to GRB 180205A. 

In this paper we present new optical photometry of the bright GRB 180205A with the COATLI telescope in the $w$ filter.
 This paper is organized as follows: in \S\ref{sec:observations}, we present our observations with COATLI and observations from other telescopes. In \S\ref{sec:analysis} we explain the afterglow model that we use, and we fit the data using segments of power-law. In \S\ref{sec:interpretation}, we interpret the observations in the context of the fireball model. Finally, in \S\ref{sec:discussion} we discuss the results and summarize our conclusions. In Appendix~\ref{app:coatli} we describe briefly the state and main features of COATLI as well as its characterization and response. 

\section{Observations}
\label{sec:observations}
~

\subsection{{\itshape Fermi} Observations}


The {\itshape Fermi}/GBM instrument triggered on GRB 180205A at 2018 February 05 04:25:25.393 UTC (trigger 539497530/180205184) and observed several pulses with a $T_{90}$ duration of $15.4\pm1.4$ seconds, making GRB 180205A a long GRB, and a 10-1000 keV fluence of $(2.1\pm0.1) \times 10^{-6}~\mathrm{erg\,cm^{-2}}$ \citep{22386,2016ApJS..223...28N}. The burst was apparently not detected by the {\itshape Fermi}/LAT instrument.

Taking into account the values of the fluence and peak energy reported by the {\itshape Fermi}/GBM collaboration \citep{22386}, the corresponding total isotropic energy is approximately $1.1\times 10^{52}$ erg  for $z=1.401$ \citep{2002A&A...390...81A}.


\subsection{Neil Gehrels Swift Observatory}

The {\itshape Swift}/BAT instrument triggered on GRB 180205A at $T =$ 2018 February 05 04:25:29.33 UTC (trigger 808625)  and observed a multi-peaked structure from about $T-7$ seconds to about $T+11$ seconds, with major peaks at about $T-5$ seconds and $T+0$ seconds \citep{22393}.

The {\itshape Swift}/XRT instrument started observing the field at $T+162$ seconds in WTSLEW mode and at $T+181$ seconds in WT mode. It detected a bright, fading source at 08:27:16.72 +11:32:31.6 J2000 with a 90\% uncertainty radius of 1.5 arcsec \citep{22381,22394}. The 0.2-10 keV flux in the initial $2.5\,{\rm s}$ image was $9.50\times 10^{-10}\,{\rm erg\, cm^{-2}\,s^{-1}}$ \citep{22381}. 
We reduced the WT and PC mode data using the pipeline described by \cite{2007AJ....133.1027B} and \cite{2007ApJ...671..656B}. We then extended the light curve to earlier times using the WTSLEW data reduced by the UK Swift Science Data Centre \citep{2009MNRAS.397.1177E} using the same counts-to-flux-density calibration as in our analysis of the WT and PC data.

The {\itshape Swift}/UVOT instrument started observing the field at $T+181$ seconds and detected a bright, fading source at 08:27:16.74 +11:32:30.9 J2000 with a 90\% uncertainty radius of 0.42 arcsec \citep{22394,22396}. The source faded in the white filter from magnitude $15.75\pm0.02$ at $T+182$ to $T+331$ seconds to magnitude $16.33\pm0.02$ at $T+540$ to $T+732$ seconds \citep{22396}.

\subsection{COATLI}
\label{sec:coatliobservations}
COATLI\footnote{\url{http://coatli.astroscu.unam.mx/}} is a robotic 50-cm telescope at the Observatorio Astron\'omico Nacional on the Sierra de San Pedro M\'artir in Baja California, Mexico \citep{2016SPIE.9908E..5OW}. COATLI is currently operating with an interim instrument, a CCD with a field of view of $12.8 \times 8.7$ arcmin and {\itshape BVRIw} filters. Our observations of GRB 180205A with COATLI are discussed in detail in Appendix \ref{app:coatli}, and are summarized briefly here.

COATLI started to observe GRB 180205A at 04:29:06.3 UTC, which was only 10.7 seconds after receiving the GCN/TAN alert packet. However, despite this fast response, our first observation started at $T+217.0$ seconds. The relatively long delay was due to a satellite telemetry problem \citep{22381}, which meant that COATLI did not receive the usual GCN/TAN BAT alert packets but only the later GCN/TAN XRT alert packet.

All of our COATLI observations are 5 second exposures in the clear $w$ filter. The read time for the CCD is about 4 seconds, so the cadence was typically about 9 seconds. The telescope dithered, taking ten images in one dither position before moving to the next dither position. Up to $T + 1500$ seconds, we consider the exposures individually. From $T + 1500$ seconds to $T + 9000$ seconds, we combine sets of 10 exposures taken over about 86 seconds. From $T + 9000$ seconds to the end of the first night we combine sets of 50 exposures taken over about 500 seconds to improve the signal-to-noise ratio. On subsequent nights, we combine all the frames with a FWHM of $2.8$ arcsec or less.

We performed aperture photometry using Sextractor \citep{1996A&AS..117..393B} with an aperture of 3.5 arcsec diameter. Table~\ref{tab:datos} gives our aperture photometry. For each image it gives the start and end times of the observation, $t_i$ and $t_f$, relative to the trigger time $T$  the total exposure time, $t_{exp}$, the AB magnitude $w$ (not corrected for Galactic extinction) and the 1$\sigma$ total uncertainty in the magnitude (including both statistical and systematic contributions).

For a wide filter like the COATLI $w$ filter, the extinction in general depends on the spectral shape (see Appendix \ref{app:coatli}).  Table~\ref{tab:beta} shows the estimated spectral indices $\beta$ using colors reported by the GROND team \citep{22383,22391} at two epochs and the corresponding values of $A_w$ for an extinction of E(B-V)=0.03 \citep{22383}. For such a low extinction the dependence on the spectral shape is weak, and so we adopt $A_w = 0.08$.

\begin{figure*}
\centering
 \includegraphics[width=0.8\textwidth]{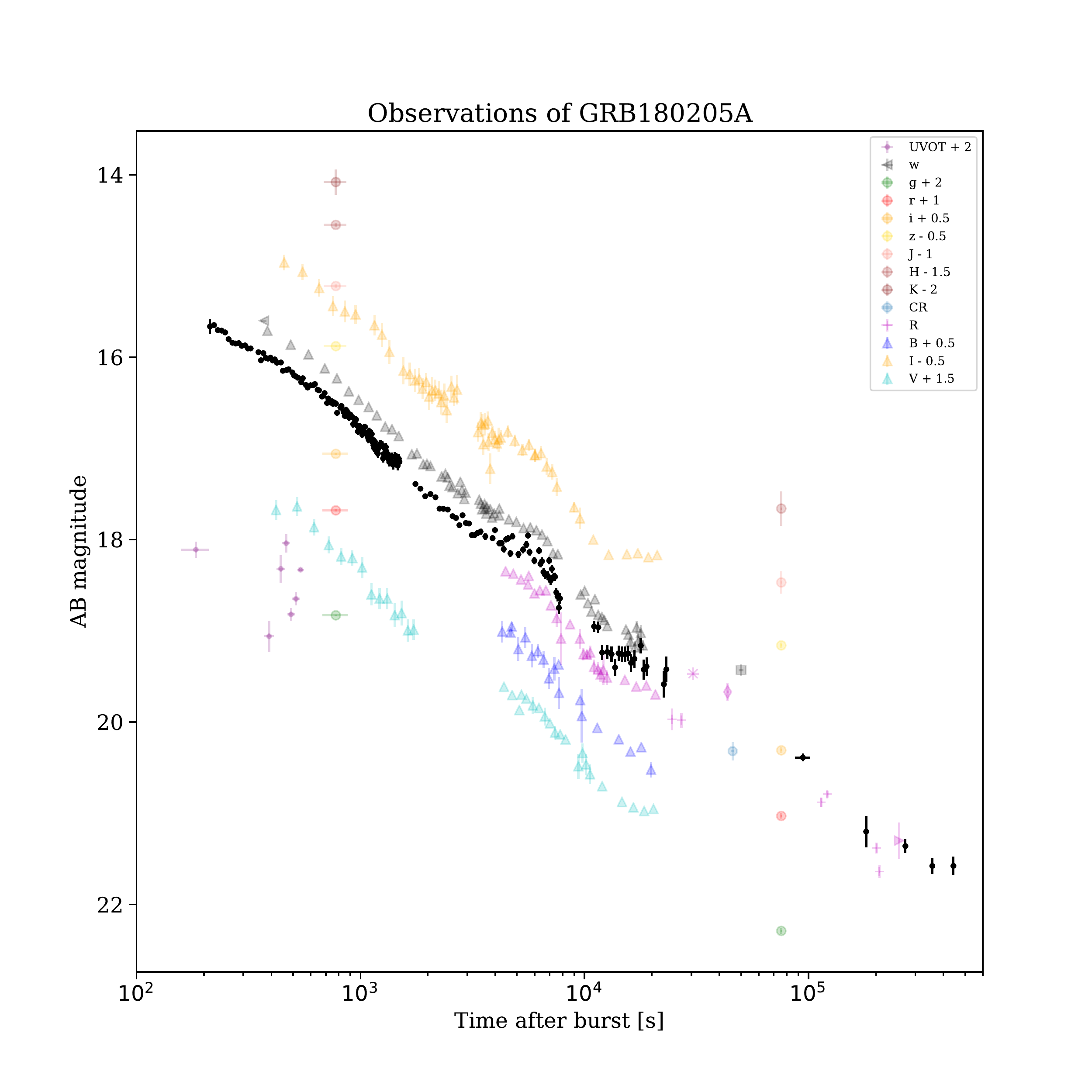}
 \caption{Light curve of GRB 180205A in different filters: {\itshape Swift}/UVOT (points) \citep{22381,22396}, GROND (circles) \citep{22383,22391}, Mondy (cross) \citep{22430}, Hankasalmi (square) \citep{22401}, Nanshan (diamonds) \citep{22395}, DOAO (plus) \citep{22404}, Nickel/KAIT (triangles) \citep{22382,22390}, and COATLI (stars) (this work). The photometry has not been corrected for Galactic extinction.}
 \label{fig:figure1}
\end{figure*}

Figure \ref{fig:figure1} shows the COATLI light curve and photometry from {\itshape Swift}/UVOT \citep{22381,22396}, GROND \citep{22383,22391}, Hankasalmi \citep{22401}, Mondy \citep{22430}, Nanshan \citep{22395}, DOAO \citep{22404} and Nickel/KAIT  \citep{22382,22390}.

\subsection{Other Terrestrial Observations}

\cite{22382} confirmed the bright, fading optical counterpart in observations with KAIT, which started observing  the field at $T+371$ seconds. They detected the source in their clear filter at about magnitude 15.6, and observed that it faded by about 1.2 magnitudes over the next 22.5 minutes. \cite{22383} reported early $grizJHK$ magnitudes obtained by GROND at about $T+13$ minutes. Further photometry was published by \cite{22391,22392,22397,22401,22404,22414}.

\cite{22384} obtained a spectrum at $T+1.5$ hours that showed many absorption features, including C~II, Si~II, Si~II*, Al~II, Fe~II, Mg~II, and Mg~I, at a common redshift of
$z=1.409$ and concluded that this was the redshift of the GRB.

\cite{22388} measured a low polarization of $P\sim0.9\%$ in $R$ at $T+2.5$ hours.
	

\section{X-ray and Optical, Temporal and Spectral analysis}
\label{sec:analysis}

\subsection{Temporal Analysis}

The prompt emission from the GRB detected by {\itshape Fermi}/GBM and {\itshape Swift}/BAT lasted until about $T + 11$ seconds. The earliest complementary data are from {\itshape Swift}/XRT starting at $T + 162$ seconds, {\itshape Swift}/UVOT at $T + 181$ seconds, and COATLI at $T + 217$ seconds. Therefore, we focus our analysis on the afterglow. In our analysis of the early afterglow, we mainly use the XRT and COATLI data, since the UVOT data are in different filters and have relatively long exposures (39-147 seconds).

\begin{figure*}
\centering
 \includegraphics[width=0.65\linewidth]{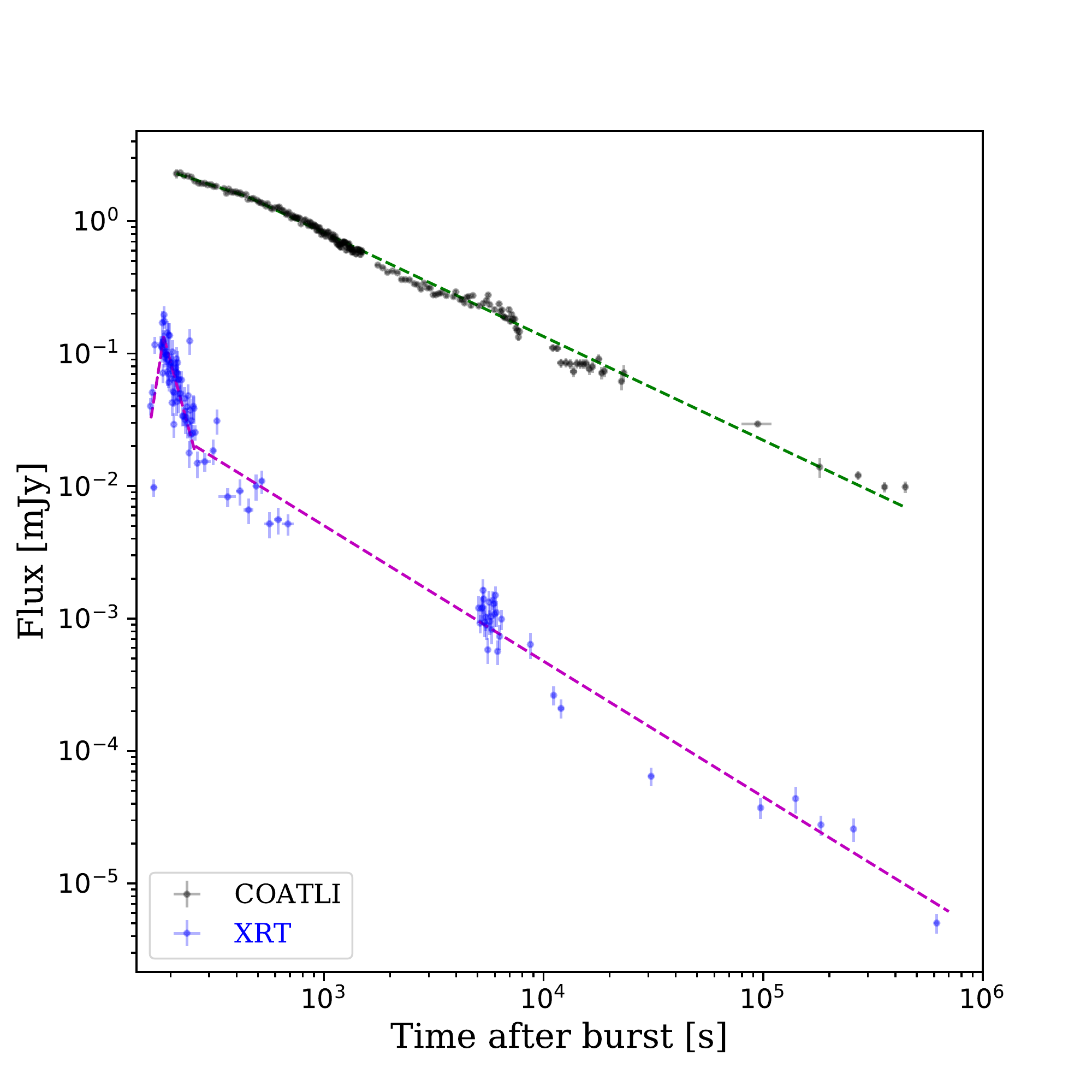}
 \includegraphics[width=0.33\linewidth]{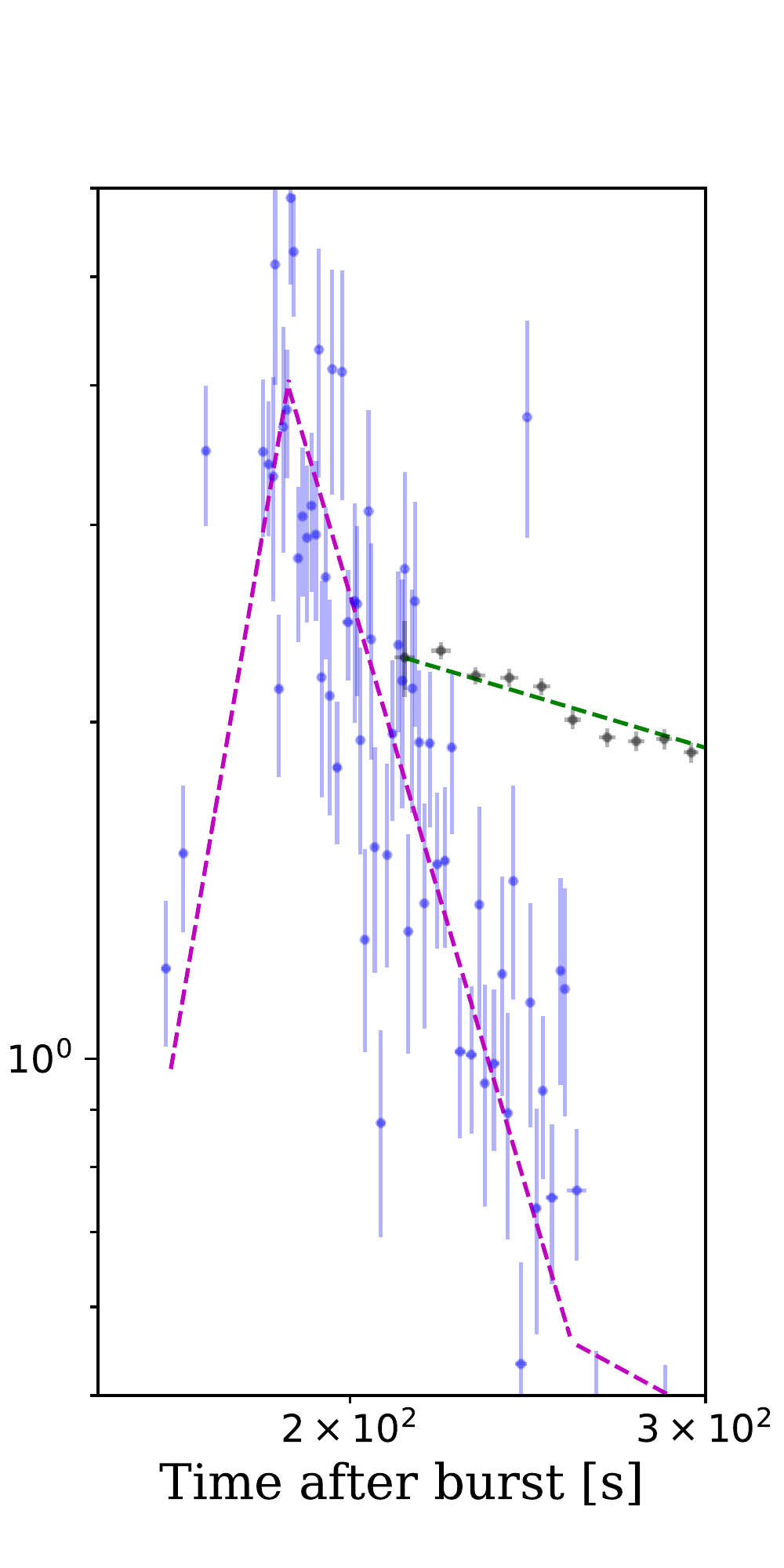}
 \caption{\emph{Left:} Light curves and broken power-law fits of GRB 180205A in $w$ from COATLI (black data and green fits) and X-rays from {\itshape Swift}/XRT (blue data and pink fits) at 1~keV. \emph{Right}: X-ray flare multiplied by a factor of 30 and the beginning of the optical plateau.}
 \label{fig:figure2}
\end{figure*}

Figure~\ref{fig:figure2} shows the optical and X-ray light curves for GRB 180205A corrected by galactic extinction.  
We will fit the light curves with segments of temporal power-laws $F_\nu \propto t^{-\alpha}$, in which $F_\nu$ is the flux density, $t$ is the time since the BAT trigger, and $\alpha$ is the temporal index. The power-law fits are summarized in Table~\ref{tab:fit}.
The relevant stages are:

\begin{enumerate}
\item The X-ray light curve for $t < 260$ seconds. In $163 < t < 260$ seconds, we see a bright flare in X-rays with a peak at about $t = 188$ seconds.  Currently, there is no widely accepted standard model for fitting an X-ray flare. Empirically, we fit two broken power-law segments in order to obtain parameters to describe it.  The rise has $\alpha_\mathrm{X,rise}=-4.82\pm1.88$ for $163 < t < 188$ seconds and the decay has $\alpha_\mathrm{X,decay}=6.08\pm1.15$ for $188 < t < 260$ seconds. We note that the uncertainty on the rising index is large due to the short time over which it is observed ($163 < t < 188$ seconds and only 7 XRT data points). Nevertheless, it is clear that the behavior prior to peak is not consistent with an extrapolation to earlier times of the subsequent decay.

\item The X-ray light curve for $t > 260$ seconds. This region can be fitted as a power-law with a temporal index of $\alpha_\mathrm{X,normal} = 0.98 \pm 0.03$. This is a typical normal decay phase seen in many afterglows \citep{2006ApJ...642..354Z}.

\item The optical light curve for $t < 454$ seconds. This region can be fitted with a temporal index of $\alpha_\mathrm{O,plateau} = 0.53 \pm 0.02$. This is a plateau phase.

Our optical observations begin at $t = 217$, significantly after the peak of the X-ray flare at $t = 188$ seconds but nevertheless before the end of the flare at $t = 260$ seconds. We see no strong evidence for a counterpart optical flare; between $t = 188$ and $t = 260$ seconds the X-ray flux falls as $F\propto t^{-6.08}$, but the optical flux only falls as $F\propto t^{-0.53}$ in accordance with the gentle power-law decline that continues to later times.

\item The optical light curve for $t > 454$. This region can be fitted as a power-law with a temporal index of $\alpha_\mathrm{O,normal} = 0.78 \pm 0.01$. This is a typical normal decay phase seen in many afterglows \citep{2006ApJ...642..354Z}.


\end{enumerate}

\begin{deluxetable*}{llcr}
\tabletypesize{\scriptsize}
\tablewidth{0pt}
\tablecaption{Temporal and Spectral Power-Law Indices\label{tab:fit}}
\tablehead{\colhead{Stage}&\colhead{Time Interval}&\colhead{Parameter}&\colhead{Index}}
\startdata

X-ray flare rise& $163<t<188$&$\alpha_\mathrm{X,rise}$&$-4.82\pm 1.88$\\
X-ray flare decay&$188<t<260$&$\alpha_\mathrm{X,decay}$&$6.08\pm1.15$\\
X-ray normal decay&$260< t$&$\alpha_\mathrm{X,normal}$&$1.02\pm0.03$\\
 &$10000< t<918651$&$\beta_\mathrm{X,normal}$&$1.13 \pm 0.10$\\[2ex]

Optical plateau& $217<t<454$&$\alpha_\mathrm{O,plateau}$&$0.53\pm0.02$\\
Optical normal decay &$454<t<500000$ &$\alpha_\mathrm{O,normal}$&$0.78\pm0.01$\\
\end{deluxetable*}

\subsection{Spectral Analysis}

We fit the photometry with a spectral power-law $F_\nu\propto \nu^{-\beta}$, in which $F_\nu$ is the flux density, $\nu$ is the frequency, and $\beta$ is the spectral index.

\begin{figure}
\centering
 \includegraphics[width=0.45\textwidth]{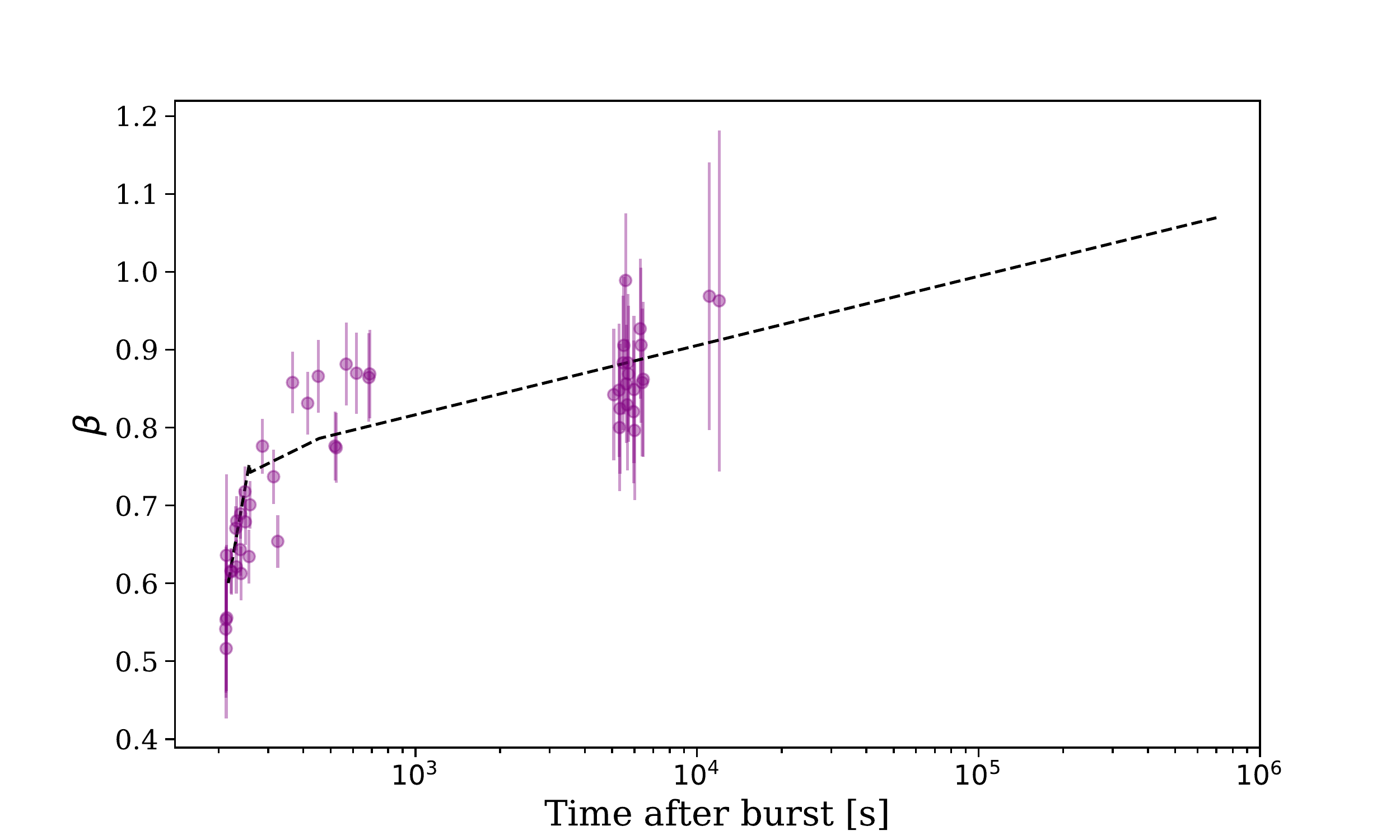}
 \caption{Temporal evolution of the spectral index $\beta$. The dashed line shows the spectral index obtained from our fits to the COATLI and XRT light curves. The points are directly calculated from the COATLI and XRT flux densities.}
  \label{fig:beta-evolution}
\end{figure}

Figure~\ref{fig:beta-evolution} shows the evolution of the SED. The dashed line shows the spectral index obtained from our fits to the COATLI and XRT light curves. The points are directly calculated from the COATLI and XRT flux densities. The spectrum evolves rapidly
from $\beta \approx 0.5$ during the flare to a steeper spectrum with $\beta \approx 0.75$ to 0.8 during the optical plateau, and then evolves more slowly to about $\beta \approx 1.0$ at late times. 

\begin{figure}
\centering
 \includegraphics[width=0.45\textwidth]{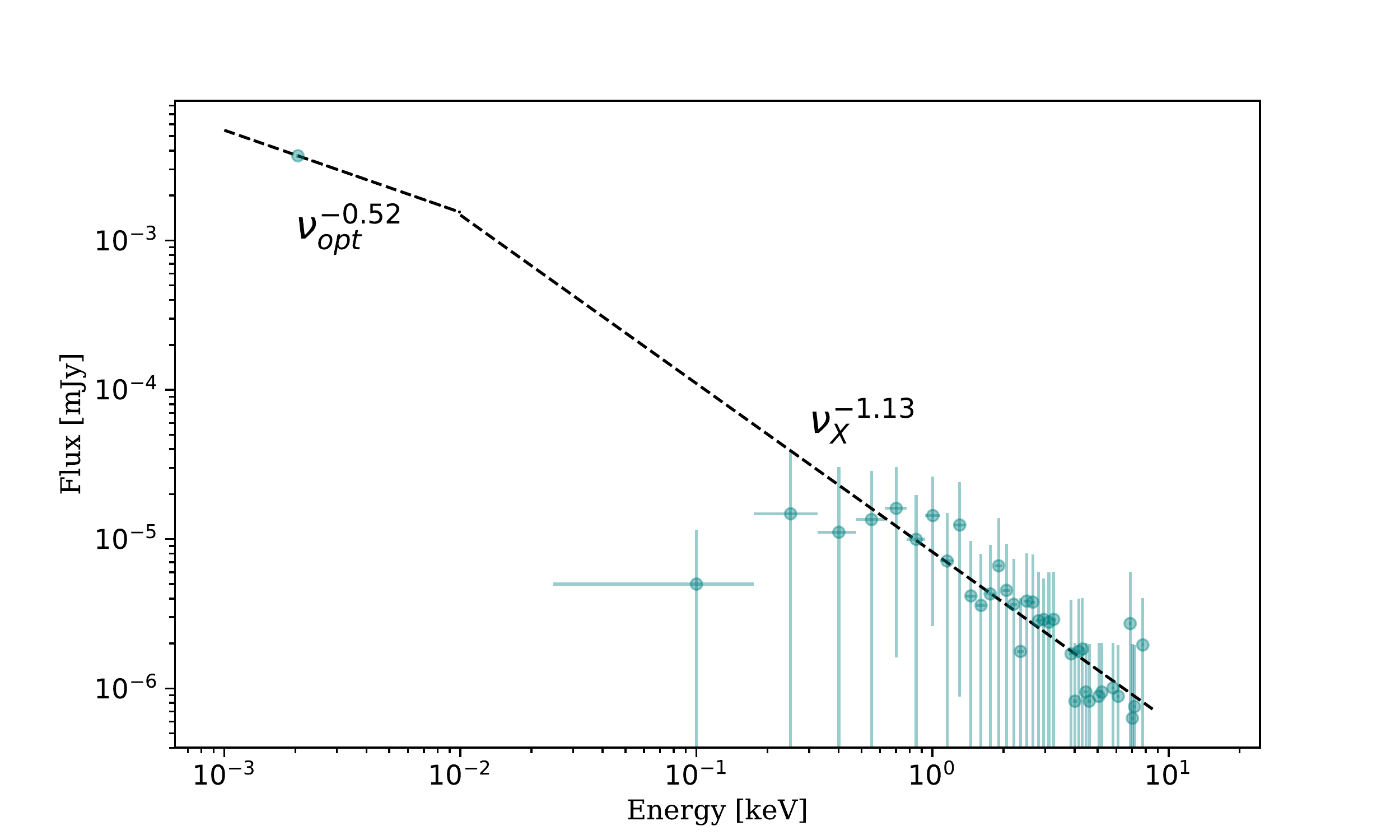}
 \caption{The SED of GRB 180205A from X-rays to the optical. The date are from COATLI and XRT in the interval $10000 < t < 918651$ seconds. The dashed line is a combination the fitted spectrum at high frequency and the predicted spectrum below the cooling break at lower frequency.}
 \label{fig:spectrum}
\end{figure}

Figure~\ref{fig:spectrum} shows the spectrum at late time in more detail. We produced the {\itshape Swift/}XRT X-ray spectrum for $10000 < t < 918651$ seconds using the pipeline described by \cite{2007AJ....133.1027B} and \cite{2007ApJ...671..656B} and added the optical point from our COATLI data interpolated over the same interval. Fitting the XRT spectrum directly, we obtain an X-ray index of $\beta_\mathrm{X,normal}=1.13\pm0.10$
, whereas at the same time the global index from the optical to X-rays is $\beta = 0.7-0.8$.
This suggests that there is a spectral break between the optical and the X-rays, with a predicted spectral index of $\beta_\mathrm{O,normal}=0.52\pm0.02$ in the optical; this is explained in more detail in the next section.

\section{Theoretical interpretation}
\label{sec:interpretation}
The long-lived emission of the afterglow is most commonly explained by the synchrotron forward-shock model. The distribution of electron energies is assumed to be given by a power-law $N(\gamma) \propto \gamma^{-p}$ in which $\gamma$ is the Lorentz factor of the electron and $p$ is the index. The observed synchrotron flux is formed by a series of power-law segments $F_{\nu}\propto t^{-\alpha}\nu^{-\beta}$ in time $t$ and frequency $\nu$. \cite{1998ApJ...3597L..17S} and \cite{2000ApJ...536..195C} derived the synchrotron light curves when the outflow is decelerated by a constant-density interstellar medium (ISM) and the stellar wind of the progenitor.  The proportionality constants of synchrotron fluxes in both regimes are explicitly computed by \cite{2016ApJ...831...22F} and \cite{2015ApJ...804..105F} for a ISM and stellar wind densities, respectively.

Assuming an efficiency of $\eta\approx$ 20\% and a typical value for \textbf{ISM} density $n=1\,{\rm cm^{-3}}$ \cite[e.g. see][]{2014ApJ...785...29S,2009ApJ...701..824N,1996ApJ...473..204S}, we derive, using the theoretical model described by \cite{2017ApJ...848...15F} and the Chi-square minimization method \citep{1997NIMPA.389...81B}, the microphysical parameters that describe the X-ray and optical afterglow of GRB 180205A. We obtain values of $\epsilon_B=8\times 10^{-4}$ and $\epsilon_e=0.1$ which agree with previous estimations of these parameters \citep[see, e.g.,][]{2010MNRAS.409..226K,2017ApJ...848...15F}.

The light curves of GRB 180205A in X-rays and in the optical are not untypical. At early times, we see an X-ray flash and a plateau in the optical. At later times, both light curves decrease more steeply in a normal decay phase. We begin by analysing the normal decay phase in order to provide a constraint on the population of relativistic electrons. We then assume continuity in this population and consider the earlier phase and possible explanations for the X-ray flare and the absence of a counterpart optical flare.

\subsection{Normal Decay Phase}
\label{sec:normal}

For $t > 454$ seconds the X-ray light curve (Figure~\ref{fig:figure2}) can be fitted with a simple power-law with a temporal index of $\alpha_\mathrm{X,normal}=1.02\pm0.03$. Furthermore, for $T+10000$ to $T+918651$ seconds the X-ray spectrum can be fitted with a simple power-law with a spectral index of $\beta_\mathrm{X,normal}=1.13\pm0.1$. We suggest that this X-ray emission arises above the cooling break ($\nu_c<\nu$) in a slow-cooling scenario \citep{1998ApJ...3597L..17S}, and so expect $F_\nu\propto\nu^{-p/2}t^{(2-3p)/4}$. Our observed value of $\alpha_\mathrm{X,normal}$ implies $p=2.03\pm 0.04$ and subsequently $\beta_\mathrm{X,normal} = 1.02 \pm 0.02$. The value of $p$ is consistent with the range of values typically for GRB afterglows and the predicted value of $\beta_\mathrm{X,normal}$ is consistent with our observation of $\beta_\mathrm{X,normal}=1.13\pm0.1$


For the optical, we suggest emission below the cooling break ($\nu_m<\nu<\nu_c$) in a slow-cooling scenario, and expect $F_\nu\propto \nu^{-(p-1)/2}t^{3(1-p)/4}$ for jet being driven into a constant-density environment \citep{1998ApJ...3597L..17S}. Our predicted value of $p$ gives a predicted value of $\alpha_\mathrm{O,normal} = 0.77 \pm 0.03$, in good agreement with the observed value of $0.78 \pm 0.01$. A jet being driven into a decreasing-density stellar-wind environment would give a much steeper decay ($1.27\pm0.03$), which is not consistent with the observed behavior. Our predicted value of $p$ also predicts a spectral index $\beta_\mathrm{O,normal} = 0.52 \pm 0.02$, which is in good agreement with the values of 0.51--0.55 observed by GROND at 780 seconds (Table~\ref{tab:beta}) but not such a good agreement with the redder values observed at later times by the same group (Table~\ref{tab:beta}). This may indicate that the cooling break is approaching the optical at later times. In Figure~\ref{fig:spectrum}, we show the combined spectrum from the optical to the X-rays, which supports the presence of a cooling break between these regimes.

It is clear that the light curves in the normal phase are not perfect power laws. For example there appear to be excesses in the optical from 5000--8000 seconds and perhaps on the second night at around 80,000 seconds. These departures in the light curve might be produced by density variations \citep{2002A&A...396L...5L}.

\subsection{Plateau Phase}

For $t < 454$ seconds, the optical light curve exhibits a flat plateau with a temporal index of $\alpha_{0,plateau}=0.53\pm0.02$. A priori, one might think that this phase corresponded to emission above the cooling break ($\nu_c < \nu < \nu_m$) in the fast-cooling scenario \citep{1998ApJ...3597L..17S}, which would give $\alpha = 0.25$. However, we are unable to find values of microphysical parameters and circumburst density that give a transition time (at which $\nu_c = \nu_m$) to the normal decay phase of around 454 seconds \citep{2017ApJ...848...94F}. Even with extreme microphyiscal conditions, this break occurs no later than $T +10$ seconds. Therefore, we reject this idea and postulate that the plateau phase is powered by late central-engine activity.

\subsection{X-Ray Flare}

While the optical light curve shows a plateau, the X-ray light curve shows a bright X-ray flare with a peak at $t = 188$ seconds and a variability timescale of $\delta t/t\sim 0.3$. The X-flare can be fitted with two power-laws with rising and decaying indices of $\alpha_{\rm X, rise}=-4.82\pm1.88$ and $\alpha_{\rm X,decay}=6.08\pm1.15$, respectively.  The X-ray flux at the peak of the flare is about 7 times larger than the X-ray flux in the normal decay phase immediately after the flare. The X-ray flare released about 5\% of the total energy observed in gamma-rays during the prompt phase. Notably, the X-ray flare does not have a counterpart in the optical. In the following subsections, we discuss the possible explanations of the origin of this flare. For the analysis, we assume the same electron index $p=2.03\pm 0.04$ derived in \S\ref{sec:normal}. 

\subsubsection{Reverse Shock Emission}

A reverse shock is expected when the expanding relativistic ejecta encounters the ISM; a forward shock is driven into the ISM and a reverse shock is driven upstream. In the reverse shock, electrons are heated and cooled primarily by synchrotron and Compton scattering emission. A reverse shock is predicted to produce a single flare \citep[see, e.g.,][]{2007ApJ...655..391K, 2018arXiv180402449F, 2012ApJ...751...33F, 2017arXiv171008514F}.  After the peak of the reverse shock, no new electrons are injected and the shell material cools adiabatically \citep{2016ApJ...818..190F}. The evolution of reverse shock can be considered in the regime where the flare is overlapped (thick) or separated (thin) from the prompt emission. Since the X-ray flare in GRB 180205A occurs much later than the prompt emission, any corresponding reverse shock must have evolved in the thin-shell case. 

\cite{2007ApJ...655..391K} discussed the generation of an X-ray flare by Compton scattering emission in the early afterglow phase when the reverse shock evolves in the thin-shell case. They found that the X-ray emission created in the reverse region displays a time variability scale of $\delta t/t\sim 1$ and  varies as $F_\nu\propto t^{5(p-1)/4}$ before the peak and $\propto t^{-(3p+1)/3}$ after the peak. However, when the high-latitude emission due to the curvature effect is present, the flux decays as $F_\nu\propto t^{-(3+p)/2}$.
For an electron spectral index of $p = 2.03\pm 0.04$, we expect the temporal index of the rise to be 1.6 and for the decay to be 2.5 in the absence of curvature effects and 2.6 in their presence.
Thus, these values cannot account for the observed values of $\alpha_{\rm X, rise}=-4.82\pm1.88$ and $\alpha_{\rm X,decay}=6.08\pm1.15$. Furthermore, as indicated by \cite{2007ApJ...655..391K}, this mechanism cannot produce the short variability timescale of $\delta t/t\approx0.3$ observed in GRB 180205A. Finally, a reverse shock would be expected to produce an optical counterpart to the X-ray flare, and this is not seen.
Therefore, we conclude that a reverse shock is not a plausible mechanism for the X-ray flare of GRB 180205A.

\subsubsection{Neutron Signature}

Another important mechanism that might explain the X-ray flare is the presence of neutrons in the fireball \citep{1999ApJ...521..640D, 2014ApJ...787..140F, 2014MNRAS.437.2187F}. At a large distance from the progenitor, when neutrons and ions have been fully decoupled, ions begin to slow down and neutrons form a leading front. Later, a rebrightening in the early afterglow is expected when neutrons decay and their daughter products in turn interact with the slowing ions \citep{2003ApJ...585L..19B}.

\cite{2005MNRAS.364L..42F} developed this mechanism analytically and predicted the light curve. They reported that when the ejecta evolve in ISM, the variability timescale should satisfy $\delta t/t\sim 1$, the flare should be present in all electromagnetic bands, and the light curve should be characterized by a slowly rising flux followed by a sharp rebrightening bump. This does not describe the flare in GRB 180205A which exhibited a variability timescale of $\delta t/t\approx0.3$, a fast rising index of $-4.82\pm1.88$, and no counterpart flare in the optical. Therefore, we conclude that neutron-proton decoupling is not a plausible mechanism for the X-ray flare of GRB 180205A.

\subsubsection{Two-Component Jet}

Ejecta with two components (one ultra relativistic and another mildly relativistic) have been proposed to explain the X-ray/optical flares and/or rebrightening emission in afterglow light curves \citep{2005ApJ...631.1022G, 2018arXiv180302978F,2008Natur.455..183R}. \cite{2005ApJ...631.1022G} derived the two-component jet light curves and found that this model cannot produce very sharp features in the light curve. In addition, the flux decay after the peak of the second component should be described by the standard afterglow model, and the variability timescale should satisfy $\delta t/t\sim 1$. Again, this model cannot reproduce the variability timescale $\delta t/t\approx 0.3$ and the fast rise and decay of the X-ray flare.

\subsubsection{Late Central-Engine Activity}

In the context of late central-engine activity, the relativistic jet contains multiple shells and the X-ray and optical flares are the result of repeated internal shocks between these shells. The light curves are the superposition of the normal prompt emission and afterglow from the initial activity and flares from late activity.

Diverse origins have been suggested for late activity, including  fragmentation of the core associated to the progenitor during the collapse \citep{2005ApJ...630L.113K}, rupture of the accretion disk due to gravitational instability at large radii in different time intervals \citep{2006MNRAS.370L..61P,2006ApJ...636L..29P} and  a neutron star with differential rotation \citep{2006Sci...311.1127D}.

The fast rise comes naturally from the short time-variability of the central engine and of the internal shocks.  The  observed variability timescale of the X-ray flare is much shorter than the duration of the late activity $\delta t/t< 1$.  For a randomly co-moving magnetic field due to internal shell collisions, the flux, in the emitting region, decays as $F_\nu\propto t^{-2 p}$ in the slow cooling regime \citep[see e.g.;][]{2006ApJ...642..354Z}.

Internal shocks induced by the late activity can account for an X-ray flare without a strong counterpart flare in the optical bands \citep[e.g.][]{2005Sci...309.1833B}.  For instance, the typical synchrotron energy is
$E^\mathrm{syn}_p=3\,\mathrm{keV}$ $\epsilon_\mathrm{e,-0.3}^2\gamma_\mathrm{ sh}\times\left(\frac{4.512}{1+z}\right)\sqrt{\gamma_\mathrm{sh}}\epsilon_\mathrm{ B,-1}^{1/2}t_\mathrm{\nu,-1}^{-1}\Gamma_\mathrm{is,3}^{-2}L_{j,52}^{1/2}$ \citep{2017ApJ...848...94F} which, for typical values of the luminosity $L$, Lorentz factor $\Gamma$, $\epsilon_B$, and $t_\nu$ is only observable in X-rays.

For GRB 180205A, assuming continuity in the value of electron spectral index  $p=2.03\pm0.04$, the X-ray flux is expected to decline rapidly with an index of $4.06 \pm 0.08$. Therefore, the value of the decaying index is in agreement with the best-fit value $\alpha_{X,decay}=6.08\pm1.15$. (We note that late activity might produce electrons with a different value of $p$ to the normal phase, but even so we do not expect a value that is much different from $p = 2.0$--2.3, and so this conclusion would remain the same.) The variability time scale of $\delta t/t\sim 0.3$ is consistent with an expected value around $\delta t/t\simeq 0.1$ for late activity.


Finally, the absence of a strong optical counterpart flare is also consistent with late activity. The total energetics of the flare suggest that late activity released about 5\% of the energy that was released during the prompt emission.

GRB 180205A is not the first event to show signatures of late central activity from early observations; there are several cases explained using this scenario, for example GRB 011121 \citep{2005MNRAS.364L..42F}, GRB 050406 \citep{2005Sci...309.1833B}, GRB 050502B \citep{2005Sci...309.1833B}, GRB 100814A 
\citep{2014A&A...562A..29N}, and GRB 161017A \citep{2018PASJ..tmp...95T}. \cite{2017MNRAS.464.4399D} studied the flares cataloged by \cite{2010MNRAS.406.2149M} and concluded that many could be explained by late activity. The flare observed in GRB 180205A is similar to many of these other flares in terms of  $t_{peak}$, duration, and $t_{decay}$, which further motivates our interpretation of its origin as being due to late central-engine activity.


\section{Summary}
\label{sec:discussion}

We have presented optical photometry of the afterglow of GRB 180205A with the COATLI telescope and interim instrument \citep{2016SPIE.9908E..5OW}. COATLI received an automated alert at $T + 206.3$ seconds, and its quick response allowed us to obtain photometry of the early afterglow from $T+217.0$ seconds, only 10.7 seconds after the alert. 


We compare the optical light curve from COATLI with the X-rays light curve from XRT. We see an early X-ray flare from the start of the observations at $T + 163$ seconds, with a peak at about $T + 188$ seconds, and lasting until about $T + 260$ seconds. The X-ray flare does not have a detectable optical counterpart flare; the early optical light curve shows a plateau to about $T + 454$ seconds. The flare and plateau are followed by a normal decay in both X-rays and the optical. 

We compare the data to the predictions of models. We find that the normal decay of the afterglow is most convincing explained as a forward shock against a surrounding homogeneous ISM rather than against a remnant stellar wind. The properties of the flare cannot be easily explained by reverse shock emission, neutron-proton decoupling, or a two-component jet. The flare is most consistently explained by late activity of the central engine; this can explain the fast rise and fall and the absence of an optical counterpart flare.

Our observations of GRB 180205A add to our understanding of the early stages of GRB evolution and the transition from the prompt phase to the early afterglow. While earlier observations give hints of the range of behavior in this transition, we think it is fair to say that observations have not yet really been able to provide strong guidance for theory. Over the next few years, we plan to use the COATLI telescope to conduct early follow-up observations of GRBs discovered by the {\itshape Neil Gehrels Swift Observatory}, and anticipate that these efforts will expand our empirical understanding of this important area.

\section*{acknowledgments}

We thank the staff of the Observatorio Astron\'omico Nacional on the Sierra de San Pedro M\'artir. We thank the anonymous referee for a very helpful report. 

Some of the data used in this paper were acquired with the COATLI telescope and interim instrument at the Observatorio Astron\'omico Nacional on the Sierra de San Pedro M\'artir. COATLI is funded by CONACyT (LN 232649, 260369, and 271117) and the Universidad Nacional Aut\'onoma de M\'exico (CIC and DGAPA/PAPIIT IT102715, IG100414, and IN109408) and is operated and maintained by the Observatorio Astron\'omico Nacional and the Instituto de Astronom{\'\i}a of the Universidad Nacional Aut\'onoma de M\'exico.

This work made use of data supplied by the UK Swift Science Data Centre at the University of Leicester.

This work made use of data from the Pan-STARRS1 DR1. The Pan-STARRS1 Surveys (PS1) and the PS1 public science archive have been made possible through contributions by the Institute for Astronomy, the University of Hawaii, the Pan-STARRS Project Office, the Max-Planck Society and its participating institutes, the Max Planck Institute for Astronomy, Heidelberg and the Max Planck Institute for Extraterrestrial Physics, Garching, The Johns Hopkins University, Durham University, the University of Edinburgh, the Queen's University Belfast, the Harvard-Smithsonian Center for Astrophysics, the Las Cumbres Observatory Global Telescope Network Incorporated, the National Central University of Taiwan, the Space Telescope Science Institute, the National Aeronautics and Space Administration under Grant No. NNX08AR22G issued through the Planetary Science Division of the NASA Science Mission Directorate, the National Science Foundation Grant No. AST-1238877, the University of Maryland, Eotvos Lorand University (ELTE), the Los Alamos National Laboratory, and the Gordon and Betty Moore Foundation.

\clearpage

\renewcommand{\thesection}{\Alph{section}}
\setcounter{section}{0}

\section{COATLI Observations}
\label{app:coatli}

\subsection{Hardware}

The COATLI telescope is an ASTELCO Systems 50-cm f/8 Ritchey-Chr\'etien reflector with protected aluminum coatings. The telescope does not currently give images better than 1.4 arcsec FWHM because of incorrectly figured mirrors; the vendor plans to replace the mirrors in fall 2018.

COATLI is equipped with an interim instrument consisting of a Finger Lakes Instruments ML3200 detector with a Kodak KAF-3200ME monochrome CCD and a Finger Lakes Instruments CFW-1-5 filter wheel. Our medium-term plans are to replace this with an imager with fast guiding and active optics \citep{2016SPIE.9908E..5OW}.

The CCD has $2184 \times 1472$ photoactive pixels each 6.8 microns or 0.35 arcsec to a side. We typically bin the CCD $2\times2$ to give 0.70 arcsec pixels. The field is $12.8 \times 8.7$ arcmin with the long axis roughly north-south. The CCD has good peak quantum efficiency from 550 to 700 nm, but is poor to the blue of 400 nm and to the red of 800 nm. The read noise is about 11 electrons.

The filter wheel has $BVRIw$ filters. The $BVRI$ filters are Bessell filters supplied by Custom Scientific. 
The $w$ filter is a anti-reflection coated BK-7 filter supplied by Custom Scientific.
All filters are 5 mm thick and 50 mm in diameter.

\begin{figure}
\centering
 \includegraphics[width=0.5\textwidth]{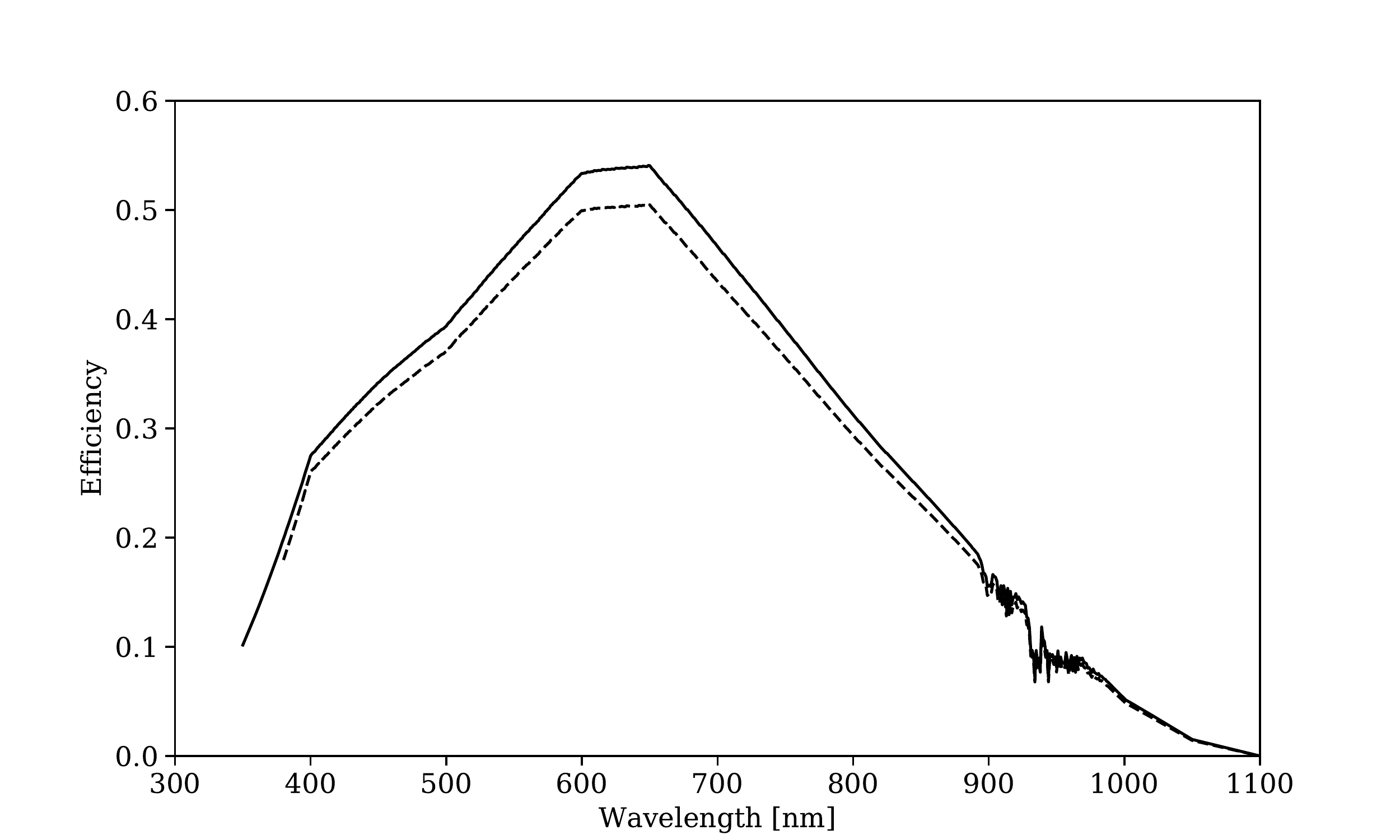}
 \caption{The estimated system efficiency of the COATLI telescope and interim imager at an airmass of 1.5. The continuous line shows the system efficiency without filters and dashed line the system efficiency with the $w$ filter.}
 \label{fig:system-efficiency}
\end{figure}

Figure~\ref{fig:system-efficiency} shows the estimated system efficiency of the COATLI telescope and interim imager at an airmass of 1.5 with and without the $w$ filter. The transmission of the $w$ filter below 380 nm is uncertain; the anti-reflection is optimized for 400-1000 nm, and we have no information on its performance below 380 nm. However, the figure shows  that even without the filter the efficiency is falling rapidly into the UV. The pivot wavelength of the system efficiency with the $w$ filter, defined by equation (A16) of \cite{2012PASP..124..140B}, is 630 nm.

The response of the COATLI $w$ filter is quite different to that of the Pan-STARRS1 $w$ filter \citep{2012ApJ...750...99T}. The rising response from 400 to 650 nm is similar, but the Pan-STARRS1 response is then flat to the sharp filter cutoff at 800 nm, whereas the COATLI response falls from 700 nm out to the limit of the CCD response at 1100 nm.

\subsection{Response to GRB 180205A}

COATLI is connected to the GCN/TAN alert system and received an XRT alert packet for GRB 180205A at 04:28:55.6 UTC ($T + 206.3$ seconds). It immediately slewed to the burst and began observing, with the first exposure starting at 04:29:02.1 UTC ($T+212.8$ seconds), only 6.5 seconds after the alert. 

This quick response is due to several factors: the telescope is mounted on an ASTELCO Systems NTM-500 mount that accelerates at up to $10~\mathrm{deg\,s^{-2}}$ to a maximum velocity of $30~\mathrm{deg\,s^{-1}}$; the control system, derived from the one used at the nearby 1.5-meter Johnson telescope with the RATIR instrument \citep{2012SPIE.8444E..5LW}, has been tuned to avoid latency; and the scheduler favors observational programs on the same side of the meridian as the center of the Swift field, which helps avoid the maneuver necessary to move from one side of the meridian to the other (since the mount is used in a German equatorial configuration).

Unfortunately, in this particular case, the rapid response was somewhat in vain. The usual GCN/TAN BAT alert packets were not sent, apparently because of a satellite telemetry problem \citep{22381}. Therefore, while the first exposure started 6.5 seconds after the alert, this corresponded to $T+212.8$ seconds after the Swift/BAT trigger.

Furthermore, there is evidence that the telescope had not settled before the first exposure started; there are streaks in the image and the raw count levels of field stars is lower than in subsequent images by a factor of 6. Therefore, we believe that the exposure effectively started at 04:29:06.3 UTC ($T + 217.0$ seconds) or 10.7 seconds after the alert and had an effective exposure time of 0.8 seconds.
 
On the first night of 2018 February 05 UTC, the telescope observed the burst almost continuously from 04:29 UTC to the end of astronomical twilight at 13:21 UTC, interrupting only to refocus about once per hour. The telescope also observed the burst on subsequent nights, finishing on 2018 February 10.

Normally, the control system takes 5 second exposures for the first half hour after the trigger and then switches to 30 second exposures, gaining sensitivity for a given total observing time. However, since the GCN/TAN XRT alert packet does not give the trigger time and since the GCN/TAN BAT alert packets were not sent, no trigger time was available and the control system persisted in using relatively inefficient short exposures. Furthermore, we did not notice this and so failed to manually switch to longer exposures on subsequent nights. (To avoid this in the future, we have now changed the control system to switch to 30 second exposures either half an hour after the trigger or half an hour after the first alert.)

\subsection{Reduction}

The raw images from the COATLI interim imager have a read noise of about 11 electrons and are sampled at 6.2 electrons per DN. This sampling is normally adequate. However, problems can arise if, for example, the sample median of a set of pixels is used as an estimator of the population median, since the sample median can differ from the population median by up to half a DN. This problem appears in Sextractor \citep{1996A&AS..117..393B}, which by default estimates the background as 2.5 times the sample median minus 1.5 times the sample mean. To avoid this problem, we add uniform noise between 0 and 1 DN to every pixel prior to reduction. The additional noise of $1/\sqrt{12}$ DN or 1.8 electrons is negligible compared to the read noise.

Our reduction pipeline performs bias subtraction, dark subtraction, flat-field correction and cosmic-ray cleaning with the \emph{cosmicrays} task in IRAF \citep{1986SPIE..627..733T}. We performed astrometric calibration of our images using the \url{astrometry.net} software \citep{2010AJ....139.1782L}.

For coadding images, we measured the offsets between the brightest star of the field and then used the  \emph{imcombine} routine of IRAF \citep{1986SPIE..627..733T}.

\subsection{Photometry}

We obtained raw aperture photometry using Sextractor \citep{1996A&AS..117..393B} with a 3.5 arcsec diameter aperture (5 pixels). We chose the aperture equal to 5 pixels after varying the aperture value between 3 and 7 pixels and observing lower noise in the light curves of the field stars. 

\begin{figure}
\centering
 \includegraphics[width=0.5\textwidth]{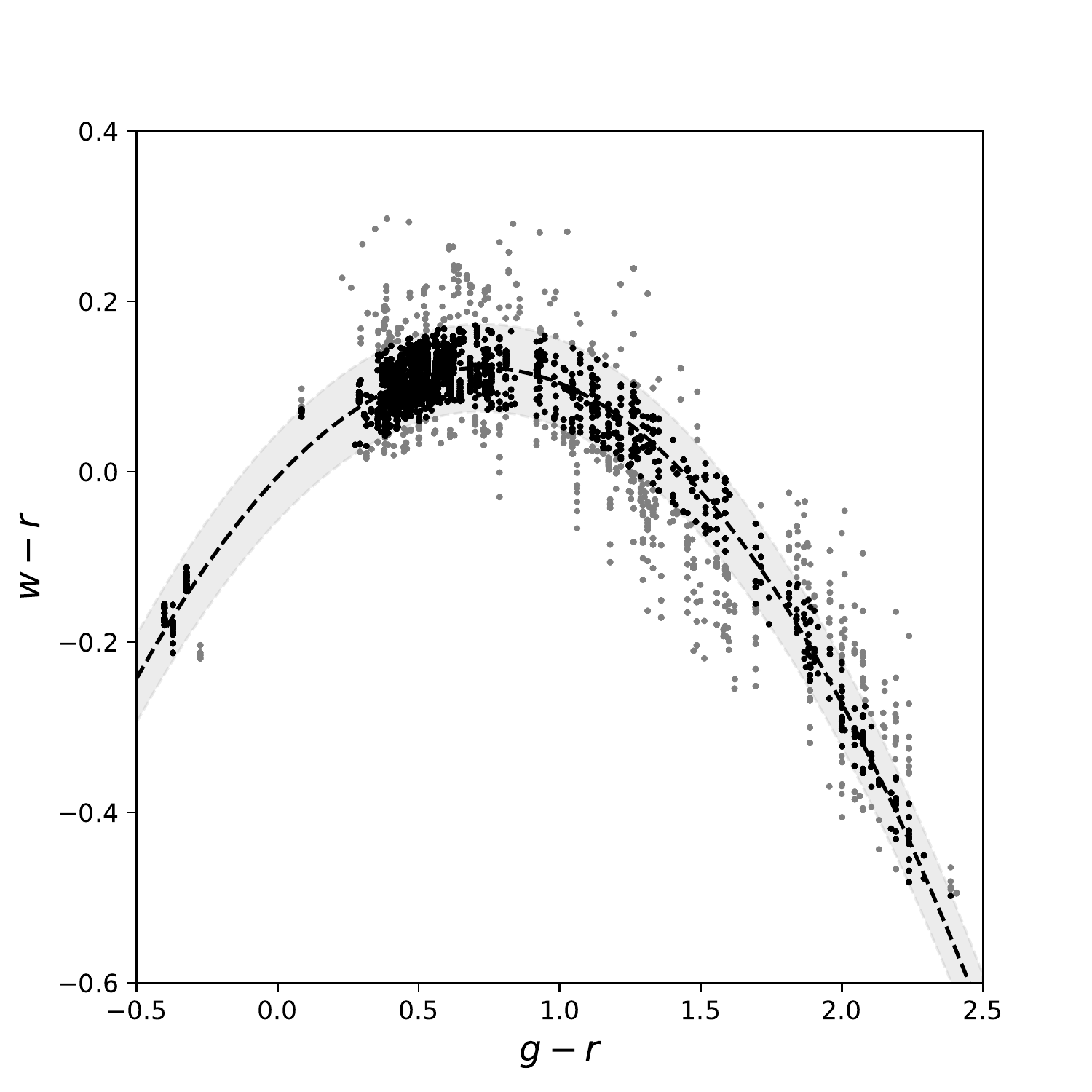}
 \caption{The transformation between \emph{w'} and Pan-STARRS1 \emph{gr}. The black solid line is the quadratic model and the grey dashed lines are $\pm3$ times the RMS deviation from the model. The black data are fitted observations. The grey data are rejected observations. The error bars are $1\sigma$.}
 \label{fig:calibration}
\end{figure}

Observations of about 3000 Pan-STARRS1 stars in other fields (Figure~\ref{fig:calibration}) show that for normal stars the transformation from Pan-STARRS1 $g$ and $r$ AB magnitudes \citep{2016arXiv161205244M} to the natural $w$ AB magnitudes of COATLI is well fitted for $-0.5 \le (g'-r') \le 2.5$ by 
\begin{equation}
(w-r')=0.353(g'-r')-0.243(g'-r')^2,
\end{equation}
with a standard deviation of $\sigma=0.017$.

We applied this equation to Pan-STARRS1 photometry of stars in the field of GRB 180205A to obtain local $w$ standards. We used these local standards to calibrate our raw photometry and produce natural $w$ AB magnitudes of the afterglow. 

\subsection{Extinction Correction}
\label{app:extinction}

The COATLI $w$ filter is wide, and so the relation between the extinction in the filter $A_w$ and the reddening $E_{B-V}$ is not trivial. Even for a fixed extinction law $A_\lambda/E_{B-V}$, the relation depends on the spectrum of the source (since blue spectra are more strongly extinguished than red spectra) and is non-linear (since extinction gradually reddens the spectrum).

Using the \cite{1989ApJ...345..245C} mean extinction curve for $R_V = 3.1$, we modeled the extinction $A_w$ as a function of the reddening $E_{B-V}$ for sources with unreddened spectra $F_\nu \propto \nu^{-\beta}$ with $-2 \le \beta \le +4$ and reddenings $0 \le E_{B-V} \le 1$. We then fitted the results, and found
\begin{equation}
A_w/E_{B-V} \approx (a_0 + a_1\beta) + (b_0 + b_1\beta)E_{B-V},
\end{equation}
in which $a_0 = +2.898$, $a_1 = -0.187$, $b_0 = -0.300$, and $b_1 = +0.026$. The deviations between the model results and the fit are at worst 2\%, and so are dwarfed by typical uncertainties in the reddening $E_{B-V}$.

\begin{deluxetable*}{ccccccc}
\tablecaption{Spectral index $\beta$ and extinction $A_w$ estimation\label{tab:beta}}
\tablehead{
\colhead{Color}&
\colhead{$\beta$ }&
\colhead{Time [s]}&
\colhead{Extinction $A_w$}}
\startdata
$g-r$&0.51&780&0.084\\
$g-i$&0.52&780&0.084\\
$r-i$&0.55&780&0.084\\
$g-r$&0.88&75600&0.082\\
$g-i$&0.93&75600&0.081\\
$r-i$&0.99&75600&0.081
\enddata
\end{deluxetable*}

\cite{1998ApJ...500..525S} estimate $E(B-V) = 0.03$ in the direction of the GRB.

\LongTables
\begin{deluxetable*}{rrrc}
\tablecaption{COATLI observations of GRB 180205A\label{tab:datos}}
\tablehead{\colhead{$t_i$ (s)}& \colhead{$t_f$ (s)}& \colhead{$t_{exp}$ (s)}&\colhead{$w$ (AB)}}

\startdata
217.0 & 217.8 & 0.8 & 15.639 $\pm$ 0.078 \\
221.8 & 226.8 & 5.0 & 15.568 $\pm$ 0.017 \\
230.8 & 235.8 & 5.0 & 15.624 $\pm$ 0.018 \\
239.8 & 244.8 & 5.0 & 15.629 $\pm$ 0.018 \\
248.8 & 253.8 & 5.0 & 15.648 $\pm$ 0.018 \\
257.8 & 262.8 & 5.0 & 15.723 $\pm$ 0.019 \\
268.1 & 273.1 & 5.0 & 15.762 $\pm$ 0.020 \\
277.2 & 282.2 & 5.0 & 15.771 $\pm$ 0.020 \\
286.2 & 291.2 & 5.0 & 15.766 $\pm$ 0.021 \\
295.2 & 300.2 & 5.0 & 15.796 $\pm$ 0.021 \\
305.3 & 310.3 & 5.0 & 15.791 $\pm$ 0.021 \\
314.3 & 319.3 & 5.0 & 15.825 $\pm$ 0.021 \\
323.3 & 328.3 & 5.0 & 15.824 $\pm$ 0.021 \\
350.4 & 355.4 & 5.0 & 15.866 $\pm$ 0.021 \\
359.3 & 364.3 & 5.0 & 15.954 $\pm$ 0.022 \\
368.3 & 373.3 & 5.0 & 15.876 $\pm$ 0.021 \\
377.4 & 382.4 & 5.0 & 15.928 $\pm$ 0.022 \\
386.4 & 391.4 & 5.0 & 15.936 $\pm$ 0.022 \\
396.4 & 401.4 & 5.0 & 15.925 $\pm$ 0.023 \\
405.4 & 410.4 & 5.0 & 15.956 $\pm$ 0.022 \\
414.4 & 419.4 & 5.0 & 15.945 $\pm$ 0.022 \\
423.4 & 428.4 & 5.0 & 15.981 $\pm$ 0.023 \\
441.4 & 446.4 & 5.0 & 15.978 $\pm$ 0.023 \\
450.4 & 455.4 & 5.0 & 16.069 $\pm$ 0.024 \\
468.4 & 473.4 & 5.0 & 16.061 $\pm$ 0.024 \\
477.4 & 482.4 & 5.0 & 16.053 $\pm$ 0.024 \\
496.6 & 501.6 & 5.0 & 16.088 $\pm$ 0.024 \\
505.6 & 510.6 & 5.0 & 16.122 $\pm$ 0.025 \\
515.1 & 520.1 & 5.0 & 16.131 $\pm$ 0.025 \\
525.5 & 530.5 & 5.0 & 16.146 $\pm$ 0.026 \\
543.6 & 548.6 & 5.0 & 16.195 $\pm$ 0.026 \\
552.6 & 557.6 & 5.0 & 16.151 $\pm$ 0.026 \\
570.6 & 575.6 & 5.0 & 16.225 $\pm$ 0.027 \\
580.8 & 585.8 & 5.0 & 16.253 $\pm$ 0.029 \\
598.8 & 603.8 & 5.0 & 16.228 $\pm$ 0.026 \\
616.8 & 621.8 & 5.0 & 16.226 $\pm$ 0.026 \\
625.8 & 630.8 & 5.0 & 16.215 $\pm$ 0.026 \\
643.8 & 648.8 & 5.0 & 16.277 $\pm$ 0.028 \\
652.8 & 657.8 & 5.0 & 16.284 $\pm$ 0.028 \\
673.2 & 678.2 & 5.0 & 16.352 $\pm$ 0.029 \\
682.2 & 687.2 & 5.0 & 16.344 $\pm$ 0.029 \\
691.2 & 696.2 & 5.0 & 16.313 $\pm$ 0.029 \\
709.2 & 714.2 & 5.0 & 16.422 $\pm$ 0.031 \\
718.2 & 723.2 & 5.0 & 16.373 $\pm$ 0.030 \\
727.2 & 732.2 & 5.0 & 16.400 $\pm$ 0.031 \\
736.2 & 741.2 & 5.0 & 16.418 $\pm$ 0.031 \\
745.2 & 750.2 & 5.0 & 16.404 $\pm$ 0.031 \\
754.2 & 759.2 & 5.0 & 16.432 $\pm$ 0.031 \\
764.1 & 769.1 & 5.0 & 16.417 $\pm$ 0.033 \\
777.0 & 782.0 & 5.0 & 16.430 $\pm$ 0.031 \\
786.0 & 791.0 & 5.0 & 16.530 $\pm$ 0.033 \\
813.1 & 818.1 & 5.0 & 16.472 $\pm$ 0.032 \\
822.2 & 827.2 & 5.0 & 16.455 $\pm$ 0.032 \\
840.2 & 845.2 & 5.0 & 16.519 $\pm$ 0.033 \\
850.5 & 855.5 & 5.0 & 16.564 $\pm$ 0.034 \\
860.7 & 865.7 & 5.0 & 16.497 $\pm$ 0.034 \\
869.7 & 874.7 & 5.0 & 16.512 $\pm$ 0.033 \\
878.7 & 883.7 & 5.0 & 16.567 $\pm$ 0.034 \\
891.1 & 896.1 & 5.0 & 16.583 $\pm$ 0.035 \\
900.1 & 905.1 & 5.0 & 16.548 $\pm$ 0.034 \\
909.1 & 914.1 & 5.0 & 16.566 $\pm$ 0.035 \\
918.1 & 923.1 & 5.0 & 16.576 $\pm$ 0.034 \\
927.1 & 932.1 & 5.0 & 16.654 $\pm$ 0.037 \\
936.1 & 941.1 & 5.0 & 16.652 $\pm$ 0.036 \\
945.1 & 950.1 & 5.0 & 16.605 $\pm$ 0.035 \\
955.2 & 960.2 & 5.0 & 16.604 $\pm$ 0.039 \\
964.2 & 969.2 & 5.0 & 16.670 $\pm$ 0.037 \\
973.2 & 978.2 & 5.0 & 16.735 $\pm$ 0.038 \\
982.2 & 987.2 & 5.0 & 16.687 $\pm$ 0.037 \\
991.2 & 996.2 & 5.0 & 16.677 $\pm$ 0.037 \\
1000.2 & 1005.2 & 5.0 & 16.715 $\pm$ 0.038 \\
1018.2 & 1023.2 & 5.0 & 16.769 $\pm$ 0.040 \\
1027.2 & 1032.2 & 5.0 & 16.728 $\pm$ 0.038 \\
1036.2 & 1041.2 & 5.0 & 16.688 $\pm$ 0.037 \\
1048.4 & 1053.4 & 5.0 & 16.685 $\pm$ 0.037 \\
1066.4 & 1071.4 & 5.0 & 16.761 $\pm$ 0.040 \\
1077.6 & 1082.6 & 5.0 & 16.794 $\pm$ 0.040 \\
1086.7 & 1091.7 & 5.0 & 16.824 $\pm$ 0.041 \\
1095.7 & 1100.7 & 5.0 & 16.732 $\pm$ 0.038 \\
1104.7 & 1109.7 & 5.0 & 16.783 $\pm$ 0.040 \\
1113.8 & 1118.8 & 5.0 & 16.828 $\pm$ 0.041 \\
1122.8 & 1127.8 & 5.0 & 16.763 $\pm$ 0.040 \\
1131.8 & 1136.8 & 5.0 & 16.821 $\pm$ 0.041 \\
1141.9 & 1146.9 & 5.0 & 16.885 $\pm$ 0.045 \\
1150.9 & 1155.9 & 5.0 & 16.917 $\pm$ 0.045 \\
1159.9 & 1164.9 & 5.0 & 16.847 $\pm$ 0.042 \\
1168.9 & 1173.9 & 5.0 & 16.934 $\pm$ 0.044 \\
1178.0 & 1183.0 & 5.0 & 16.912 $\pm$ 0.044 \\
1187.0 & 1192.0 & 5.0 & 16.962 $\pm$ 0.046 \\
1196.0 & 1201.0 & 5.0 & 16.976 $\pm$ 0.046 \\
1205.0 & 1210.0 & 5.0 & 16.919 $\pm$ 0.044 \\
1214.0 & 1219.0 & 5.0 & 16.892 $\pm$ 0.043 \\
1223.0 & 1228.0 & 5.0 & 16.873 $\pm$ 0.043 \\
1233.0 & 1238.0 & 5.0 & 16.873 $\pm$ 0.047 \\
1242.0 & 1247.0 & 5.0 & 16.878 $\pm$ 0.043 \\
1251.0 & 1256.0 & 5.0 & 16.885 $\pm$ 0.043 \\
1261.7 & 1266.7 & 5.0 & 17.029 $\pm$ 0.048 \\
1270.7 & 1275.7 & 5.0 & 16.981 $\pm$ 0.047 \\
1279.7 & 1284.7 & 5.0 & 16.917 $\pm$ 0.045 \\
1288.7 & 1293.7 & 5.0 & 16.926 $\pm$ 0.045 \\
1297.7 & 1302.7 & 5.0 & 16.905 $\pm$ 0.045 \\
1306.7 & 1311.7 & 5.0 & 16.994 $\pm$ 0.046 \\
1315.8 & 1320.8 & 5.0 & 17.009 $\pm$ 0.048 \\
1325.9 & 1330.9 & 5.0 & 16.983 $\pm$ 0.050 \\
1334.9 & 1339.9 & 5.0 & 17.007 $\pm$ 0.047 \\
1343.9 & 1348.9 & 5.0 & 17.069 $\pm$ 0.050 \\
1352.9 & 1357.9 & 5.0 & 17.028 $\pm$ 0.049 \\
1361.9 & 1366.9 & 5.0 & 17.060 $\pm$ 0.050 \\
1381.2 & 1386.2 & 5.0 & 17.016 $\pm$ 0.049 \\
1399.2 & 1404.2 & 5.0 & 17.100 $\pm$ 0.052 \\
1418.3 & 1423.3 & 5.0 & 17.011 $\pm$ 0.049 \\
1427.3 & 1432.3 & 5.0 & 17.072 $\pm$ 0.050 \\
1436.3 & 1441.3 & 5.0 & 17.028 $\pm$ 0.048 \\
1447.1 & 1452.1 & 5.0 & 17.019 $\pm$ 0.048 \\
1465.1 & 1470.1 & 5.0 & 17.110 $\pm$ 0.052 \\
1474.2 & 1479.2 & 5.0 & 17.054 $\pm$ 0.050 \\
1483.1 & 1488.1 & 5.0 & 17.034 $\pm$ 0.049 \\
1492.1 & 1497.1 & 5.0 & 17.068 $\pm$ 0.050 \\
1806.7 & 1889.7 & 50.0 & 17.362 $\pm$ 0.024 \\
1898.7 & 1986.7 & 50.0 & 17.445 $\pm$ 0.025 \\
1995.7 & 2106.7 & 50.0 & 17.421 $\pm$ 0.025 \\
2115.7 & 2199.7 & 50.0 & 17.456 $\pm$ 0.026 \\
2208.7 & 2291.7 & 50.0 & 17.581 $\pm$ 0.028 \\
2300.7 & 2382.7 & 50.0 & 17.583 $\pm$ 0.029 \\
2391.7 & 2513.7 & 50.0 & 17.591 $\pm$ 0.028 \\
2522.7 & 2614.7 & 50.0 & 17.663 $\pm$ 0.030 \\
2624.7 & 2707.7 & 50.0 & 17.683 $\pm$ 0.029 \\
2719.7 & 2800.7 & 50.0 & 17.762 $\pm$ 0.030 \\
2810.7 & 2891.7 & 50.0 & 17.653 $\pm$ 0.029 \\
2901.7 & 2995.7 & 50.0 & 17.738 $\pm$ 0.032 \\
3004.7 & 3087.7 & 50.0 & 17.744 $\pm$ 0.032 \\
3096.7 & 3178.7 & 50.0 & 17.870 $\pm$ 0.034 \\
3187.7 & 3271.7 & 50.0 & 17.870 $\pm$ 0.033 \\
3290.7 & 3381.7 & 50.0 & 17.846 $\pm$ 0.033 \\
3390.7 & 3472.7 & 50.0 & 17.832 $\pm$ 0.033 \\
3481.7 & 3833.7 & 50.0 & 17.885 $\pm$ 0.036 \\
3842.7 & 3925.7 & 50.0 & 17.905 $\pm$ 0.034 \\
3934.7 & 4017.7 & 50.0 & 17.815 $\pm$ 0.034 \\
4118.7 & 4212.7 & 50.0 & 17.961 $\pm$ 0.036 \\
4221.7 & 4303.7 & 50.0 & 17.959 $\pm$ 0.036 \\
4313.7 & 4398.7 & 50.0 & 18.025 $\pm$ 0.039 \\
4407.7 & 4511.7 & 50.0 & 17.917 $\pm$ 0.035 \\
4520.7 & 4605.7 & 50.0 & 17.906 $\pm$ 0.034 \\
4623.7 & 4705.7 & 50.0 & 18.071 $\pm$ 0.039 \\
4714.7 & 4807.7 & 50.0 & 17.886 $\pm$ 0.034 \\
5024.7 & 5118.7 & 50.0 & 18.081 $\pm$ 0.038 \\
5270.7 & 5353.7 & 50.0 & 18.032 $\pm$ 0.038 \\
5455.7 & 5538.7 & 50.0 & 17.976 $\pm$ 0.037 \\
5547.7 & 5629.7 & 50.0 & 17.876 $\pm$ 0.034 \\
5638.7 & 5720.7 & 50.0 & 18.057 $\pm$ 0.038 \\
5913.7 & 6023.7 & 50.0 & 18.150 $\pm$ 0.041 \\
6224.7 & 6305.7 & 50.0 & 18.043 $\pm$ 0.037 \\
6315.7 & 6400.7 & 50.0 & 18.185 $\pm$ 0.041 \\
6414.7 & 6506.7 & 50.0 & 18.157 $\pm$ 0.043 \\
6515.7 & 6597.7 & 50.0 & 18.276 $\pm$ 0.046 \\
6606.7 & 6701.7 & 50.0 & 18.302 $\pm$ 0.045 \\
6802.7 & 6902.7 & 50.0 & 18.317 $\pm$ 0.048 \\
6911.7 & 6997.7 & 50.0 & 18.150 $\pm$ 0.042 \\
7006.7 & 7093.7 & 50.0 & 18.369 $\pm$ 0.046 \\
7102.7 & 7197.7 & 50.0 & 18.242 $\pm$ 0.043 \\
7206.7 & 7335.7 & 50.0 & 18.329 $\pm$ 0.047 \\
7345.7 & 7426.7 & 50.0 & 18.328 $\pm$ 0.046 \\
7436.7 & 7525.7 & 50.0 & 18.499 $\pm$ 0.052 \\
7536.7 & 7628.7 & 50.0 & 18.544 $\pm$ 0.056 \\
7637.7 & 7719.7 & 50.0 & 18.668 $\pm$ 0.063 \\
7728.7 & 7811.7 & 50.0 & 18.565 $\pm$ 0.058 \\
10806.7 & 11270.7 & 250.0 & 18.871 $\pm$ 0.064 \\
11279.7 & 11758.7 & 250.0 & 18.881 $\pm$ 0.062 \\
11767.7 & 12221.7 & 250.0 & 19.158 $\pm$ 0.080 \\
12230.7 & 12946.7 & 250.0 & 19.152 $\pm$ 0.080 \\
12957.7 & 13447.7 & 250.0 & 19.177 $\pm$ 0.083 \\
13456.7 & 13934.7 & 250.0 & 19.321 $\pm$ 0.091 \\
13943.7 & 14438.7 & 250.0 & 19.167 $\pm$ 0.083 \\
14447.7 & 14939.7 & 250.0 & 19.176 $\pm$ 0.087 \\
14949.7 & 15401.7 & 250.0 & 19.177 $\pm$ 0.087 \\
15412.7 & 15893.7 & 250.0 & 19.169 $\pm$ 0.086 \\
15902.7 & 16386.7 & 250.0 & 19.272 $\pm$ 0.096 \\
16395.7 & 17053.7 & 250.0 & 19.227 $\pm$ 0.099 \\
17577.7 & 18071.7 & 250.0 & 19.083 $\pm$ 0.087 \\
18080.7 & 18648.7 & 250.0 & 19.346 $\pm$ 0.108 \\
18657.7 & 19226.7 & 250.0 & 19.311 $\pm$ 0.099 \\
22264.7 & 22916.7 & 250.0 & 19.505 $\pm$ 0.146 \\
22925.7 & 23495.7 & 250.0 & 19.343 $\pm$ 0.143 \\
78412.8 & 110409.6 & 29230.0 & 20.309 $\pm$ 0.042 \\
164722.3 & 197031.0 & 3400.0 & 21.122 $\pm$ 0.171 \\
251164.0 & 290421.4 & 19175.0 & 21.280 $\pm$ 0.076 \\
337589.2 & 376417.5 & 20035.0 & 21.498 $\pm$ 0.091 \\
424008.9 & 463199.5 & 18930.0 & 21.497 $\pm$ 0.101 \\

\enddata
\end{deluxetable*}

\end{document}